\documentclass[twocolumn,reprint,superscriptaddress,amsmath,amssymb,aps,nofootinbib]{revtex4-1}

\usepackage{graphicx}
\usepackage{dcolumn}
\usepackage{bm}
\usepackage{hyperref}
\usepackage[utf8]{inputenc}
\usepackage[dvipsnames]{xcolor}
\usepackage[normalem]{ulem}
\usepackage{xspace}
\usepackage{fontawesome} 
\usepackage{multirow}
\usepackage{footmisc}
\usepackage{tabularx}

\newcommand\myshade{80}
\colorlet{mylinkcolor}{ForestGreen}
\colorlet{mycitecolor}{Red}
\colorlet{myurlcolor}{violet}

\hypersetup{
  linkcolor  = mylinkcolor!\myshade!black,
  citecolor  = mycitecolor!\myshade!black,
  urlcolor   = myurlcolor!\myshade!black,
  colorlinks = true
}

\newcommand{\eV}{\text{e\kern-0.2ex V}\xspace}

\newcommand{\GeV}{\text{G\eV}\xspace}
\newcommand{\TeV}{\text{T\kern-0.1ex \eV}\xspace}

\definecolor{tobycolour}{rgb}{.5,.0,.5}

\newcommand{\GRAPPA}{%
Gravitation Astroparticle Physics Amsterdam (GRAPPA),\\
Institute for Theoretical Physics Amsterdam
and Delta Institute for Theoretical Physics,\\
University of Amsterdam, Science Park 904, 1098 XH Amsterdam, The Netherlands}
\newcommand{\TRIUMF}{%
TRIUMF, 4004 Wesbrook Mall, Vancouver, BC V6T 2A3, Canada}
\newcommand{\yale}{Department of Astronomy \& Department of Physics, Yale University, New Haven, CT 06511, USA}
\newcommand{\ucb}{Berkeley Center for Theoretical Physics, University of California, Berkeley, CA 94720, USA}
\newcommand{\ucla}{Department of Physics and Astronomy, University of California, 
Los Angeles, CA 90095-1547, USA}
\newcommand{\oregon}{Department of Physics, University of Oregon, Eugene, OR 97403 USA}
\newcommand{\cern}{Theoretical Physics Department, CERN, 1211 Geneva, Switzerland}
\newcommand{\jhu}{Department of Physics and Astronomy, Johns Hopkins University, Baltimore, MD 21218, USA}
\newcommand{\ncsu}{North Carolina State University, Department of Physics, Raleigh, North Carolina 27695-8202, USA}
\newcommand{\RICE}{Department of Physics and Astronomy, Rice University,\\
1600 Main Street, Houston, TX 77025, USA}

\begin{document}
\title{Gravitational wave probes of dark matter: challenges and opportunities}

\author{Gianfranco Bertone}%
\email{g.bertone@uva.nl}
\affiliation{\GRAPPA}%
\author{Djuna Croon}%
\email{dcroon@triumf.ca}
\affiliation{\TRIUMF}
\author{Mustafa A. Amin}
\email{mustafa.a.amin@rice.edu}
\affiliation{\RICE}
\author{Kimberly K. Boddy}%
\email{kboddy@jhu.edu}
\affiliation{\jhu}
\author{Bradley J. Kavanagh}
\email{b.j.kavanagh@uva.nl}
\affiliation{\GRAPPA}
\author{Katherine J. Mack}
\email{kmack@ncsu.edu}
\affiliation{\ncsu}
\author{Priyamvada Natarajan}
\email{priyamvada.natarajan@yale.edu}
\affiliation{\yale}
\author{Toby Opferkuch}
\email{toby.opferkuch@cern.ch}
\affiliation{\cern}
\author{Katelin Schutz}
\email{kschutz@berkeley.edu}
\affiliation{\ucb}
\author{Volodymyr Takhistov}
\email{vtakhist@physics.ucla.edu}
\affiliation{\ucla}
\author{Christoph Weniger}
\email{c.weniger@uva.nl}
\affiliation{\GRAPPA}
\author{Tien-Tien Yu}
\email{tientien@uoregon.edu}
\affiliation{\oregon}

\begin{abstract}
In this white paper, we discuss the prospects for characterizing and identifying dark matter using gravitational waves, covering a wide range of dark matter candidate types and signals. We argue that present and upcoming gravitational wave probes offer unprecedented opportunities for unraveling the nature of dark matter and we identify the most urgent challenges and open problems with the aim of encouraging a strong community effort at the interface between these two exciting fields of research.
\end{abstract}

\maketitle

\section{Introduction}

The direct detection of gravitational waves (GWs) has opened a new era and window into the study of the Universe \cite{Abbott:2016blz,TheLIGOScientific:2017qsa,GBM:2017lvd,LIGOScientific:2018mvr}, allowing us to probe in exquisite detail the nature of compact astrophysical objects and their environments. In the coming decades, a range of current and future experiments will continue to study GWs, including LIGO/Virgo \cite{LIGOScientific:2019vkc}, LISA \cite{AmaroSeoane:2012km,2017arXiv170200786A}, the Einstein Telescope \cite{Sathyaprakash:2012jk}, Pulsar Timing Arrays \cite{2010CQGra..27h4013H,2013CQGra..30v4009K}, and others. These experiments promise to address, and possibly solve, a variety of longstanding problems in astrophysics, cosmology, and particle physics (see e.g.~Ref.~\cite{Barack:2018yly} and references therein). 


Here, we present a survey of the most promising prospects for characterizing dark matter (DM) using gravitational wave detectors. Exploiting synergies between these two very active fields of research, we review the status of six different areas of research. We show that tremendous advances in our understanding are likely using this combined approach. Our paper is organized around the following topics, which are also summarized in Fig.~\ref{fig:fig1}: 

\begin{figure*}[tbh!]
\centering
   \includegraphics[width=\linewidth]{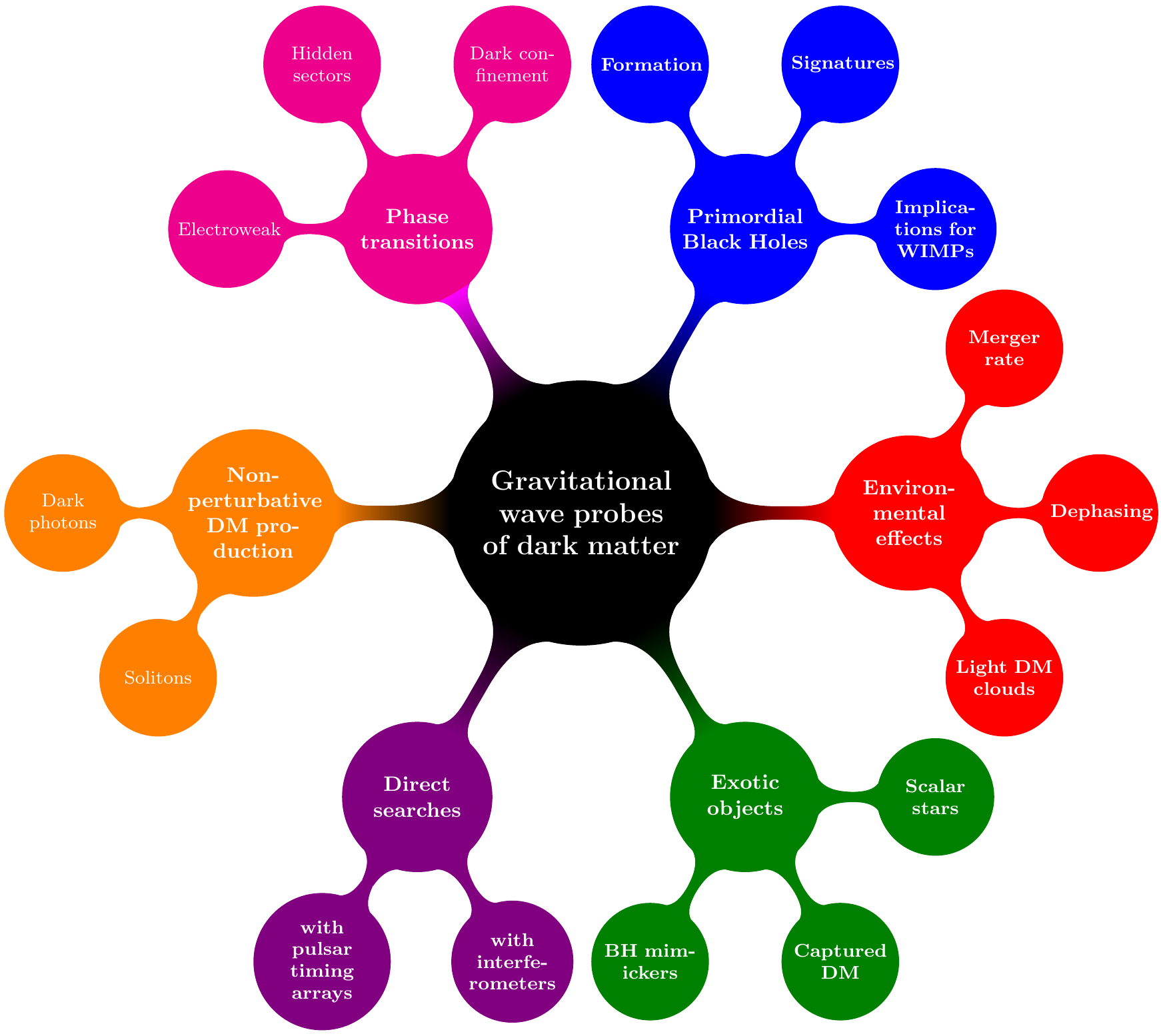}
   \caption{Mind map of gravitational wave probes of dark matter. See text for details.
   }
   \label{fig:fig1}
\end{figure*}

\begin{itemize}
    \item {\bf Primordial black holes.} In Sec.~\ref{sec:PHBs}, we discuss the connections between DM and primordial black holes (PBHs). We briefly review the currently proposed PBH formation scenarios, the strategies to discriminate this population from astrophysical black holes using GWs and other probes, and the implications that their discovery would have for particle DM.
    
    \item {\bf Environmental effects.} In Sec.~\ref{sec:env}, we review the so-called ``environmental'' effects induced by DM around compact objects -- such as the de-phasing of gravitational waveforms and the spin-down of black holes due to super-radiance -- and the prospects for probing them with GW experiments. 
    
    \item{{\bf Exotic compact objects.} In Sec.~\ref{sec:exo}}, we discuss the possibility that DM forms compact objects, or that it accretes onto and modifies astrophysical objects, and explore the ensuing gravitational wave signatures.
    
    \item {\bf Direct detection with GW experiments.} In Sec.~\ref{sec:direct}, we review ideas on how to repurpose GW experiments (inteferometers and Pulsar Timing Arrays) to  search for the signature of DM not via gravitational wave signals, but ``directly,'' e.g.~through the effects induced by DM particle physics couplings, or time-varying gravitational potentials. 
    
    \item  {\bf Non-perturbative DM dynamics.} In Sec.~\ref{sec:NP}, we consider scenarios in which DM particles are produced non-perturbatively and discuss the detectability of the stochastic gravitational wave background that they would produce. 
    
    \item {\bf Phase transitions.} Finally, in Sec.~\ref{sec:PT}, we review the prospects for detecting a stochastic GW background induced by cosmological phase transitions, and the implications that this detection would have for DM. 
    
\end{itemize}

For each of these areas, we identify the most urgent challenges, open problems and the prospects for resolution for these outstanding questions. We argue that present and upcoming gravitational wave probes offer unprecedented opportunities for DM studies, and we encourage a strong community effort at the interface between these two exciting fields of research.

\section{Primordial Black Holes}
\label{sec:PHBs}
Primordial black holes (PBHs) have been of longstanding theoretical interest~\cite{zel1967hypothesis,Carr:1974nx}, particularly as a possible component of DM~\cite{chapline1975cosmological}. 
Such interest has been reinvigorated with the dawn of GW astrophysics, as PBHs have the potential to produce signals for current and future GW experiments~\cite{Nakamura:1997sm,Clesse:2015wea,bird2016did}. The existence of non-baryonic matter (i.e.~matter that does not appreciably interact with the electroweak or strong sectors of the Standard Model) can be seen as early as Big Bang Nucleosynthesis~\cite{Cyburt:2015mya}, indicating that in order for PBHs to be a viable DM candidate they must have formed less than $\sim$1 second after the Big Bang. PBHs with masses below $\sim 10^{-16}$~M$_\odot$ cannot constitute a substantial fraction of the DM, because they are not stable to Hawking evaporation on cosmological timescales and this can drastically change cosmological observables~\cite{Carr:2009jm}. Meanwhile, the existence of DM dominated $\sim10^6$~M$_\odot$ dwarf galaxies like Segue 2~\cite{Kirby:2013isa} indicate that PBHs cannot constitute a large fraction of the DM above this mass scale. Between these limiting masses, there are many constraints on the fraction of DM that can be comprised of PBHs, $f_\text{PBH}$: these limits on $f_\text{PBH}$ arise from constraints on extragalactic gamma rays~\cite{Carr:2009jm}; gravitational microlensing~\cite{Tisserand:2006zx,Novati:2013fxa,Niikura:2017zjd} and lensing of type Ia supernovae~\cite{Zumalacarregui:2017qqd}; the survival of star clusters in dwarf galaxies~\cite{Brandt:2016aco} and of wide stellar binaries~\cite{Monroy-Rodriguez:2014ula}; and the cosmic microwave background~\cite{Ali-Haimoud:2016mbv,Poulin:2017bwe}. For the remainder of this section, we focus on the possibility of detecting PBHs via their GW signatures and the implications of such a discovery for particle DM candidates.

\subsection{Formation}
\label{sec:PBHformation}
PBHs have a number of formation scenarios, each of which could leave slightly different GW signatures. Large curvature perturbations can be generated in hybrid~\cite{GarciaBellido:1996qt,Lyth:2010zq} and axion-curvaton~\cite{Kawasaki:2012wr} models of inflation, which can lead to the formation of PBHs. Phase transitions in the early universe can lead to topological defects such as cosmic strings, loops of which can collapse into PBHs~\cite{polnarev1991formation,MacGibbon:1997pu} (and the dynamics of the loops may be a potential source of gravitational radiation). PBHs may also be formed by the fragmentation of extended solitonic and pseudo-solitonic field configurations~\cite{Cotner:2016cvr,Cotner:2017tir,Cotner:2018vug}, or as a byproduct of the metastability of the electroweak vacuum~\cite{Kawasaki:2016ijp,Espinosa:2017sgp}. These scenarios can possibly be distinguished by the shape of the inferred PBH mass spectrum, which impacts any potential GW signal from PBH merger events. Further, PBHs formed in a matter-dominated era (e.g.~\cite{Cotner:2016cvr,Cotner:2017tir,Cotner:2018vug}) are expected to possess large spins~\cite{Harada:2017fjm}, unlike PBHs formed in the radiation-era as with standard inflationary perturbations; the LIGO Collaboration has placed some limits on the spin distribution of binary black holes through the merger events detected in Observing Runs 1 and 2~\cite{LIGOScientific:2018jsj}.
Additionally, some formation scenarios are accompanied by a tensor mode~\cite{Matarrese:1997ay}, which impacts the gravitational wave background and can thus be probed with pulsar timing and interferometric searches~\cite{Saito:2008jc,2011PhRvD..83h3521B,Orlofsky:2016vbd,Inomata:2016rbd,Bartolo:2018evs,Cai:2018dig}.

\subsection{Detection} 



 GWs have the potential to provide evidence for PBHs in the near future, as there are at least two detection scenarios that appear incompatible with black holes of astrophysical origin. The first scenario is the detection of nearby mergers of sub-solar mass BHs with LIGO/Virgo: according to the standard theory of stellar evolution, black holes do not form below the Chandrasekhar mass of $\sim 1.4\,M_\odot$, and therefore the detection of the merger of sub-solar mass BHs using LIGO and Virgo would strongly suggest a primordial origin (though see Refs.~\cite{Shandera:2018xkn,Kouvaris:2018wnh} for alternative exotic formation mechanisms). The second detection scenario is the observation of BH-BH mergers at high redshift with next generation interferometers such as Einstein Telescope (ET) \cite{Hild:2010id} and Cosmic Explorer (CE) \cite{Sathyaprakash:2012jk}: above a redshift of $z \sim 40$, the merger rate of astrophysical BHs should be negligible \cite{Koushiappas:2017kqm}. Very high-redshift mergers would therefore point towards a primordial BH population which formed much earlier \cite{Chen:2019irf}.

 PBHs may also be detected by studying populations of merger events. PBH binaries are expected to form abundantly before matter-radiation equality \cite{Nakamura:1997sm,Ioka:1998nz,Sasaki:2016jop} (while binary formation in the late Universe is sub-dominant \cite{Bird:2016dcv,Ali-Haimoud:2017rtz}). The resulting merger rate of these binaries has been studied extensively (see, for example,  Refs.~\cite{Hayasaki:2009ug,Raidal:2017mfl,Chen:2018czv,Ballesteros:2018swv,Belotsky:2018wph,Kavanagh:2018ggo,Bringmann:2018mxj,Raidal:2018bbj}). Comparing the merger rate histories of primordial and astrophysical BHs, it should therefore be possible to distinguish between them (see e.g.~\cite{Sasaki:2018dmp} for a recent review).

Other strategies to discriminate between primordial and astrophysical BHs include: the analysis of their eccentricities \cite{Cholis:2016kqi}, mass function \cite{Kovetz:2017rvv}, as well as angular momentum distributions \cite{Fernandez:2019kyb,Arbey:2019jmj}, which are already being constrained via BH-BH merger waveforms~\cite{LIGOScientific:2018jsj}; the study of the spatial distribution and mass function of radio and X-ray sources powered by the accretion of gas onto Galactic BHs ~\cite{Fender:2013ei, Gaggero:2016dpq,2018arXiv181207967M}; and the search for a population of compact objects with a large optical depth to GW lensing \cite{Jung:2017flg}.

\subsection{Implications for particle DM}
\label{sec:implications}

If PBHs do not make up all of the DM, they are generically expected to accrete a dense mini-halo of DM. This begins before matter-radiation equality, as the sphere of gravitational influence of each PBH grows \cite{Mack:2006gz,Ricotti:2007jk}. Analytic considerations suggest that such mini-halos would have a very steep central density profile $\rho(r) \propto r^{-9/4}$ \cite{Bertschinger:1985pd}, which has recently been confirmed in numerical simulations \cite{Gosenca:2017ybi,Adamek:2019gns}. The growth of the halo is expected to continue at least until PBHs are subsumed into bound structures, at which point the DM halo may be 100 times more massive than the PBH itself \cite{Mack:2006gz}. 

Such large over-densities would have profound implications for particle DM. Enhanced DM annihilation within mini-halos would potentially make solar-mass PBHs bright $\gamma$-ray sources \cite{Bertoni:2015mla,Hooper:2016cld,Schoonenberg:2016aml}, as well as contributing to the diffuse $\gamma$-ray background \cite{Taylor:2002zd,Ullio:2002pj,Ando:2005hr}. By comparing with $\gamma$-ray observations,
the abundance of PBHs can be constrained below the level of $f_\mathrm{PBH} \lesssim 10^{-8}$, assuming that the DM is a Weakly Interacting Massive Particle (WIMP) produced as a thermal relic \cite{Lacki:2010zf,Eroshenko:2016yve,Boucenna:2017ghj,Adamek:2019gns}. This suggests that PBHs and WIMPs are fundamentally incompatible. More recently, it has been pointed out that if PBHs are detected in near-future GW searches (as discussed above), this would point towards an abundance greater than $f_\mathrm{PBH} \gtrsim 10^{-5}$ \cite{Bertone:2019vsk}. This in turn would place stringent constraints on models of WIMP DM, ruling out thermal WIMP DM with GeV-scale masses and above. Such a detection would also strongly constrain large regions of the parameter space for models such as the minimal supersymmetric standard model, even when they predict only a sub-dominant population of WIMPs.

The existence of PBHs may also have more indirect implications.  
In scenarios where PBHs are formed from enhanced primordial perturbations (Sec.~\ref{sec:PBHformation}), the same rare, large density fluctuations which produce PBHs should also lead to the formation of gravitationally bound ultra-compact mini-halos (UCMHs) of particle DM \cite{Kolb:1994fi,Dokuchaev:2002fm,Berezinsky:2013fxa,Gosenca:2017ybi,Nakama:2019htb}. This in turn would lead to enhanced lensing and annihilation signatures due to these UCMHs (see e.g.~\cite{Ricotti:2009bs,Bringmann:2011ut}). Finally, the detection of PBHs would shed light on a number of models that suggest a common origin for PBHs and particle DM
\cite{Hasegawa:2017jtk,Hooper:2019gtx}.

\section{Environmental Effects}
\label{sec:env}


Future GW experiments should allow for precise reconstructions of the properties of inspiraling and merging compact objects, including mass measurements at the sub-percent level \cite{Seoane:2013qna}. With such exquisite precision, these experiments would be sensitive to tiny deviations in the gravitational waveforms which may be induced by matter in the violent environments of the merging BHs and NSs \cite{Barausse:2014tra,Barausse:2014pra}. These ``environmental effects,'' once detected, could perhaps signal the presence of DM, although similar effects due to baryons must also be taken into account.

\subsection{Cold DM}
\label{sec:environment_CDM}

As discussed above, large overdensities of cold DM may form around primordial black holes. Numerical and analytical studies of the mergers of ``dressed'' black holes show that the distribution of DM around them dramatically affects the dynamical evolution of the binaries \cite{Kavanagh:2018ggo}.

Cold DM overdensities are also expected to inevitably form around astrophysical black holes, although the slope and normalization of the DM density profile depends strongly on the formation mechanism. This process has been explored in the context of supermassive black holes lying at the centers of galaxies \cite{Gondolo:1999ef,Gondolo:2000pn,Ullio:2001fb,Bertone:2001jv,Merritt:2002vj,Bertone:2005hw,Merritt:2006mt}, as well as for intermediate-mass BHs \cite{Bertone:2005xz,Zhao:2005zr,Bringmann:2009ip}. 

The dynamical friction induced by DM particles is expected to modify the dynamics of the merger, possibly leaving an imprint on the gravitational waveform, in the form of a change in phase relative to the inspiral without DM \cite{Macedo:2013qea,Barausse:2014tra,Barausse:2014pra,Eda:2013gg,Eda:2014kra,Barack:2018yly}. Much remains however to be understood about the evolution of these systems, as only approximate solutions have been obtained thus far. The full problem of evolving the BH-BH pair -- including post-Newtonian corrections as well as the gravitational feedback induced on the DM distribution -- has yet to be solved. We also stress that the dephasing induced by a distribution of DM might substantially alter the waveform, to the point that dedicated templates would be needed in order to extract the predicted signal from the noise.

If this dephasing effect is observed, it would potentially allow for the detection of DM around black holes, as well as a measurement of the DM density.
It has been suggested that these ``dark dresses'' are incompatible with light bosonic and fermionic DM and with self-annihilating DM \cite{2019arXiv190611845H} and
their observation may therefore hint at the nature of the DM particle.

\subsection{Ultralight bosons}
\label{sec:ULBs}
Another attractive class of DM models consists of ultralight bosonic particles such as the axion, axion-like particles (ALPs), and dark photons. For DM candidates with masses below $\sim{\rm keV}$, the occupation numbers are high enough that one can describe the DM as a classical field. These light bosonic fields can form gravitational bound states around spinning black holes, and subsequently extract energy and angular momentum from the BH through a process known as ``rotational superradiance” (see~\cite{Brito:2015oca} for a review). Notably, this process depends only on the gravitational interaction between the bosonic fields and the black hole.\footnote{Superradiance can also occur around rotating neutron stars if there is some non-gravitational interaction between the bosonic cloud and the neutron star~\cite{Cardoso:2017kgn,Day:2019bbh}, although it remains to be understood if the gravitational-wave emission is relevant in this case.} Superradiance drives an exponential growth of bosonic clouds around the BH, potentially reaching masses of up to $\sim 10\%$ of the BH, which causes the BH to spin-down~\cite{Arvanitaki:2010sy,Dolan:2012yt,Brito:2014wla,East:2017ovw}. Therefore, the existence of these bosonic fields can be probed indirectly by measuring the masses and spins of BHs.  Since this process is most efficient when the Compton wavelength of the bosonic field is of comparable size to the BH, observations of stellar mass black holes (supermassive black holes) set limits on the scalar masses that apply in the range $\sim10^{-13}-10^{-11}~\rm{eV}$ ($10^{-18}-10^{-16}~\rm{eV}$)~\cite{Arvanitaki:2014wva,Brito:2017zvb}. The superradiance process is more effective for vector particles, and in that case the limits shift to $\sim10^{-14}-10^{-11}~\rm{eV}$ ($10^{-20}-10^{-17}~\rm{eV}$), for stellar mass (supermassive) black holes, respectively~\cite{Baryakhtar:2017ngi}. 

In addition, the oscillations of the bosonic clouds can source GWs that are detectable with LIGO, LISA, and other gravitational wave detectors~\cite{Arvanitaki:2014wva,Baryakhtar:2017ngi,Brito:2017zvb}. There are three types of signals that can arise from the bosonic cloud: graviton emission from level transitions, boson annihilations into gravitons, and a bosenova collapse of the boson cloud~\cite{Arvanitaki:2010sy}. The first two processes result in monochromatic GWs. The frequency of the signal is determined by the masses of the boson and black hole. For stellar mass black holes, this corresponds to the frequencies probed by Advanced LIGO while supermassive black holes correspond to the frequencies of LISA. The third process is a consequence of the self-interactions; if the attractive self-interactions are stronger than the gravitational binding energy, the bosonic cloud collapses, resulting in a burst of GWs. The amplitude of these waves is smaller, but may be observable for supermassive black holes. The gravitational wave signatures are also altered if the black hole is a part of a binary system. Most notably, the existence of a companion black hole induces resonant mixing between the growing and decaying modes of the bosonic cloud~\cite{Baumann:2018vus,Hannuksela:2018izj}. Some consequences on the gravitational wave signal are a Doppler modulation of the frequency, as well as modifications to the waveforms from the cloud's multipole moments and tidal deformations caused by the companion. 

DM in the form of ALPs may also give rise to electromagnetic signatures around compact objects, creating a new opportunity for multimessenger astronomy. For example, dense clouds of ALPs (grown by superradiance around rotating BHs) may lead to stimulated photon emission. This gives rise to a periodic radio signal, relevant for axion masses above $\sim$$10^{-8}\,\mathrm{eV}$ and BH masses below $\sim$$0.01\,M_\odot$ \cite{Rosa:2017ury,Ikeda:2019fvj}. Another possibility is that ALPs may be converted into radio photons via the Primakoff effect \cite{Primakoff:1951pj}. This process requires strong magnetic fields, making the magnetospheres of neutrons stars (NS) a promising environment for this axion-photon conversion \cite{Huang:2018lxq,Hook:2018iia,Safdi:2018oeu}. Around intermediate mass black holes (IMBHs), the DM density is expected to be significantly enhanced (see Sec.~\ref{sec:implications} and Sec.~\ref{sec:environment_CDM}) leading to an even larger predicted signal in NS-IMBH binaries. GW observations of such a system could provide a measurement of the DM density (through the dephasing effect), fixing the normalization of the expected radio signal \cite{Edwards:2019tzf}. Joint  observations of these systems using LISA and the Square Kilometer Array \cite{Bull:2018lat} would thus probe the natural parameter space of the QCD axion in the mass range $m_a \in [10^{-7}, 10^{-5}] \,\mathrm{eV}$.

\subsection{Baryons} 



Baryons can also play a major role in driving the coalescence of Massive Black Hole Binaries (MBHBs), whether in the form of stars or gas. The dominant mechanism for driving a change in the orbital dynamics of MBHBs has yet to be fully understood, but several channels are possible. Stars that are ejected from the binary loss cone via gravitational slingshot can lead to hardening of the binary, as the ejected star carries away energy~\cite{mikkola1992evolution,Rajagopal:1994zj}. This process will stall if the loss cone is depleted, but the loss cone can be efficiently replenished in sufficiently non-spherical galaxies~\cite{2011ApJ...732L..26P,2013ApJ...773..100K,2015ApJ...810...49V}. Gas disks can also play a role via tidal torquing of the binary if the mass of the disk is similar to the mass of the secondary BH, although this is complicated by star formation in the disk~\cite{2011MNRAS.417L..66N,2009MNRAS.398.1392L}. If the binary accretes infalling molecular clouds, this can also efficiently drive MBHBs into the regime of gravitational wave domination~\cite{2017MNRAS.472..514G}. 

Baryons can have a residual effect on the MBHB even after it has entered the GW-dominated phase. For instance, while the emission of GWs tends to circularize elliptical orbits, interactions with a circumbinary gaseous disk~\cite{2011MNRAS.415.3033R} or stars~\cite{sesana2013insights,2010ApJ...719..851S} can increase the orbital eccentricity, particularly for binaries with a large mass discrepancy. Such effects may be a fingerprint for baryonic influence on the binaries, but may also make it harder to identify the subtle environmental effects expected from Dark Matter.
\section{Exotic Binary Mergers}
\label{sec:exo}
If dark particles coalesce into exotic compact objects (ECOs) of astrophysical size, they may form new binary systems. The gravitational radiation emitted when such binary systems merge may be observed in gravitational wave experiments. 
Binary black hole mergers are often separated into three sequential phases: the inspiral, the merger, and the ringdown. Gravitational wave emission builds up in the inspiral phase and peaks during the merger. The end of the inspiral phase is characterized by the innermost stable circular orbit (ISCO), which can be defined for ECOs in analogy with black holes, \cite{Giudice:2016zpa}
\begin{equation}\label{eq:isco}
f_{\rm ISCO} = \frac{C_*^{3/2}}{3^{3/2}\pi G_N (M_1 + M_2)}
\end{equation}
where $M_{1,2}$ are the masses of the stars in the binary and $C_* = G_N M_*/R_*$ is the typical compactness of an ECO of mass $M_*$ and radius $R_*$, limited from above by black holes ($C_* = 1/2$). ECOs with $C_* >4/9$ are in violation of Buchdahl's theorem \cite{Buchdahl:1959zz,Cardoso:2019rvt}\footnote{With certain caveats, Buchdahl's theorem states that the compactness of a spherically symmetric self-gravitating object composed of a spherically symmetric perfect fluid cannot be arbitrarily close to that of a BH.}; we will describe some examples in Section \ref{sec:BHmimic}.
The best detection prospect for an ECO merger is for $f_\mathrm{ISCO}$ within the frequency window of an experiment. This implies that ground-based interferometers are most sensitive to $10<f_{\rm ISCO}/{\rm Hz} < 10^3$, which includes BNS, BBH, and solar mass-sized ECOs, whereas PTAs are sensitive to supermassive BHs and other compact objects with $M_* \sim 10^6 \, M_{\odot}$.

The experimental prospects also rely on the formation history and abundance of the ECOs. ECOs may form at redshifts long before first-star formation, and depending on the fraction of DM that they constitute, may not follow an NFW profile. ECOs with masses close to solar masses are also constrainable via the non-observation of microlensing events.

\subsection{Exotic stars}
Real and complex scalar fields provide excellent DM candidates; these fields can support ultra-compact, coherent solitonic configurations held together by gravity  (oscillatons \& boson stars) \cite{PhysRevLett.66.1659,Liebling:2012fv}, self-interactions (oscillons \& Q-balls) \cite{Kusenko:1997si,Copeland:1995fq,LEE1992251,Nugaev:2019vru}, or both. 
When gravity is the dominant force holding the solitons together, their mass is given by $M\lesssim 0.6 m_{\rm pl}^2/m_\phi$ \cite{PhysRev.172.1331} where $m_\phi$ is the mass of the scalar. Their compactness $C_*$ can be comparable to that of Neutron stars ($C_*\sim 0.1$).  
For sufficiently compact solitons, and a small mass $m_\phi$ ($M\sim m_{\rm pl}^2/m_\phi$), 
such objects can source detectable GWs for LIGO/LISA via their mergers with one another, or with other compact objects (see for example, \cite{Palenzuela:2006wp,Palenzuela:2017kcg,Helfer:2018vtq,Bezares:2018qwa,Clough:2018exo}).  

Boson stars stabilized by a repulsive self-interaction can form compact, astronomically sized Bose-Einstein condensates with mass $M_* = \mathcal{O}(10^{-2})\, \sqrt{\lambda_\phi} M_p^3/m_\phi^2$ \cite{Colpi:1986ye}. The maximum compactness of such  spherically symmetric stable stars is $C_* \lesssim 0.16$ \cite{Croon:2018ybs}. 
In this case, it is seen that the binary merger signal peaks at $f_{\rm peak} \sim 10^{-14} (m_\phi/{\rm eV})^2/\sqrt{\lambda_\phi} \,{\rm Hz}$, which falls within the LISA window for $m_\phi \sim \lambda_\phi^{1/4} \times {\rm MeV} $
\cite{Croon:2018ftb}. 

The differences in tidal deformability of scalar stars \cite{Cardoso:2017cfl,Sennett:2017etc,Guo:2019sns} and the possibility of energy loss through scalar radiation provide avenues for distinguishing their gravitational wave signatures from those of neutron stars and black holes. In certain cases, the gravitational wave output from the mergers of such scalar stars can exceed that of their corresponding mass black-hole counterparts \cite{Helfer:2018vtq,Bezares:2018qwa}. 

Finally, asymmetric DM comprised of fermions can also form stable compact objects \cite{Kouvaris:2015rea,Gresham:2018rqo}. Typically, self interactions are required to allow for efficient formation, see for example Ref.~\cite{Chang:2018bgx} for a recent discussion. The resulting fermion stars, due to Fermi repulsion, are in general less compact compared to boson stars in the absence of additional strong attractive interactions. As for the case of boson stars these scenarios can be distinguished through differences in tidal deformability \cite{Maselli:2017vfi}. 

\subsection{Captured DM}
Small (sub-Chandrasekhar-mass) PBHs constituting a significant fraction of DM could be efficiently captured in DM-rich environments by neutron stars (or white dwarfs). A captured black hole, growing within the star, will eventually consume the host. This can lead to novel multi-messenger coincidence signatures with GWs (e.g. a kilonova without a binary merger GW counterpart) as well as the appearance of binaries with solar-mass black holes that typically do not arise in standard astrophysics~\cite{Fuller:2017uyd,Takhistov:2017nmt,Takhistov:2017bpt}.

Alternatively, neutron star properties can be modified in the presence of particle DM. This can lead to observable deviations in the BNS mergers probed by both current and future GW observatories. Effects considered thus far in the literature include modifications to the tidal deformability through the presence of DM in the NS core \cite{Panotopoulos:2017idn,Ellis:2018bkr} as well as by extended DM clouds around coalescing NSs \cite{Nelson:2018xtr}, axionic induced fifth-forces \cite{Hook:2017psm,Huang:2018pbu} and long-range dark forces affecting the inspiral phase \cite{Croon:2017zcu,Sagunski:2017nzb,Kopp:2018jom,Alexander:2018qzg,Fabbrichesi:2019ema,Choi:2018axi}. In addition, many of these effects will also be present in extreme-mass-ratio inspirals (EMRIs), where the duration of the waveform will be observable for significantly longer periods of time in comparison to LIGO/VIRGO NS-NS binaries.

\subsection{Black hole mimickers}\label{sec:BHmimic}
Several proposed horizonless compact objects mimic the space-time of black holes at large distances as well as in the vicinity of the photon ring. Such objects are referred to as clean-photon sphere objects (ClePhOs).
ClePhOs violate the Buchdahl limit by violating one or several of its axioms \cite{Cardoso:2019rvt}, such as in alternative gravitational theories \cite{Buoninfante:2019swn}. 
Abundant ClePhOs may form a sub-fraction of DM, and will be subject to the same constraints as PBHs; we will describe a few examples here. 
\emph{Gravastars} \cite{Giddings:1992hh} are supported by negative pressure from radiative QFT effects in curved space-times, and do not necessarily require new physics. Stable gravastars may exist for a range in compactness \cite{Camilo:2018goy}. \emph{Wormholes} \cite{Einstein:1935tc} connect different regions of space-time. Solutions with different geometries exist, and can have any mass or compactness. Their stability and formation depends on the theory of (modified) gravity, though generically they are unstable \cite{Gonzalez:2008wd}. \emph{Anisotropic stars} are ECOs subject to large anisotropic stresses. Although covariant studies of anisotropic stars are challenging, it is believed they can exist for a wide range in mass and compactness \cite{Raposo:2018rjn}.

ClePhOs may partake in binary mergers, and can be distinguished from black holes through their nonzero tidal Love numbers \cite{Porto:2016zng}, which leads to higher order post-Newtonian corrections (tidal heating at 2.5PN, tidal deformability at 5PN) \cite{Maselli:2017cmm}. EMRIs hold a unique potential to probe black hole mimickers, as any nonzero tidal Love number of the central object leads to large post-Newtonian deviations of the gravitational waveform in the long inspiral phase, creating a clear signature of a non-standard BH-BH inspiral scenario \cite{Pani:2019cyc}.
\section{Direct DM detection with GW experiments}
\label{sec:direct}

\subsection{Searches with interferometers}

Cold DM particles with sub-eV masses feature in general large
occupation numbers of low-momentum states.  This is a consequence of the high
number densities required to yield the observed DM energy density.
Sub-eV CDM is hence typically described in terms of classical fields rather
than distinct particles (as already discussed in Sec.~\ref{sec:ULBs}).  For DM masses around $10^{-13}$ to $10^{-12}$ eV,
the DM field oscillation frequencies match the best sensitivity of LIGO ($\sim 100$
Hz). The DM fields are expected to be in coherent oscillation over length scales
of $\ell_\text{coh.} \sim 10^6\,\text{km}$ in the Milky Way, and might exhibit
topological defects depending on the details of the production
mechanism~\cite{Vilenkin1985-qh}. A wide class of sub-eV CDM models
exist, and, as we will discuss below, the optical cavities of gravitational wave detectors have a unique
potential to provide complementary probes to previously unconstrained parts of
the sub-eV DM parameter space. However, there are also situations where significantly heavier DM can be constrained using interferometers. For instance, if DM is composite with a super-Planckian mass (\textit{e.g.}, if DM is made of ``dark blobs'') and interacts with baryons via a long-range mediator, it can induce an observable acceleration on interferometer elements~\cite{Grabowska:2018lnd}. Furthermore, Ref.~\cite{Cheng:2019vwy} recently proposed optical cavities as sensitive probes to detect the Brownian motion caused by interacting electroweak-scale DM particles.

\textit{Dark photons.}  Ultra-light dark photons, produced via the
mis-alignment mechanism, can constitute all of dark
matter~\cite{Nelson2011-mv, Graham:2015rva,Agrawal:2018vin}.  If this dark photon DM (DPDM) is
associated with the $U(1)_B$ or $U(1)_{B-L}$ symmetries, DPDM couples directly
to baryon or neutron number and hence acts as a fifth force.  The strongest
constraints on such scenarios come from tests of the weak equivalence
principle~\cite{Su1994-jv, Schlamminger2007-gd}.  In Ref.~\cite{Pierce:2018xmy}
it was shown that DPDM with masses around $10^{-13}$ to $10^{-12}\,\text{eV}$
can potentially lead to GW signatures that are similar to monochromatic
stochastic GWs.  Since the coherence length of DPDM in this
mass-range is much larger than the separation between various GW detectors on
Earth (or the future LISA detector), the cross-correlation between measurements
from individual GW detectors can be used significantly enhance the
signal-to-noise ratio of the otherwise stochastic signal.  The first searches for
DPDM, using LIGO’s first observing run from the detectors in Hanford and
Livingston, O1, have been conducted~\cite{Guo2019-xf}, and the constraints obtained already exceed
those of fifth force searches for DM masses $m \sim 10^{-14} -
10^{-13}\,\text{eV}$.  Both methodological improvements of the analysis
technique and additional data have the potential to further strengthen
constraints on the coupling  parameter $\epsilon^2$ by more than an order of
magnitude. On the other hand, the future LISA detector will probe DM masses
around $m \sim 10^{-17}\,\text{eV}$.

\textit{Axions.}  The presence of axion DM affects the propagation of
photons by inducing minuscule changes in the phase velocity of circularly
polarized light.  The authors of Ref.~\cite{Nagano:2019rbw} (see also Refs.~\cite{DeRocco:2018jwe, Liu:2018icu, Obata:2018vvr}) have proposed a
new scheme for exploiting and enhancing this effect in the linear optical
cavities of gravitational wave detection experiments.  The basic idea is to
inject a linearly polarized laser beam into the optical cavities and search for
the generation of orthorgonally polarized light, which would be a signature of
axion DM.  The detection strategy is thought to be resilient against
the most common systematics that plague GW detectors, since they would affect
both circular polarizations in similar ways.  Projected sensitivies indicate
that this search strategy can probe axion masses below around
$10^{-11}\,\text{eV}$---up to three orders of magnitude below the current
helioscope bound from the CAST experiment~\cite{Zioutas2005-pb,
CAST_Collaboration2017-zm}.

\textit{General searches.} Ref.~\cite{Grote2019-vv} (see also Ref.~\cite{Morisaki:2018htj}) studies the effects of
bosonic sub-eV DM fields coupling to the photon kinetic or the fermion
mass terms.  Coherent oscillations of dark matter fields or the collisional
encounters of topological defects induce spatial and temporal variations of
physical constants.  Recently, various clock-clock comparison experiments have
been conducted to search for time varying physical constants~\cite{Hees2016-vx,
Van_Tilburg2015-mu}.  These experiments are mainly sensitive to sub-Hz
frequencies, with DM masses $m \lesssim 10^{-15}\,\text{eV}$.  Searches with the
cryogenic resonant-mass detector AURIGA~\cite{Branca2017-hx} provide the most
sensitive constraints on a narrow frequency around a kHz.  In the optical
cavities of GW detectors, the variation of physical constants leads (like in
the case of DPDM) to additional forces on the freely suspended mirrors, as well
as variations of the extent of the beam splitter (through variations of the
Bohr radius) and its optical index.  Ref.~\cite{Grote2019-vv} shows that
(Fabry-Perot-)Michelson interferometers, like GEO 600 and Advanced LIGO (for
DM masses $m = 10^{-13} - 10^{-11}\,\text{eV}$) or the future LISA (for $m =
10^{-16} - 10^{-18}\,\text{eV}$), have the potential to probe scalar DM models
in previously unconstrained regions of the parameter space. Minor modifications
of the experimental configurations (e.g.\ changing the thickness of
Fabry-Perot mirrors in one of the arms) can lead to significant enhancements of
the DM reach.

\subsection{Searches with PTAs}
Millisecond pulsars are excellent clocks, emitting radiation at regular intervals over long periods of time.
Given their stability, these objects are ideal for searches of GWs in the nHz frequency range~\cite{Detweiler:1979wn,Sazhin}.
The presence of GWs causes variations in the time of arrival of the electromagnetic pulses, and worldwide efforts are underway to detect the stochastic GW background using pulsar timing arrays (PTAs).
Notably, the International Pulsar Timing Array (IPTA)~\cite{Verbiest:2016vem} is a consortium of three collaborations: the North American Nanohertz Observatory for gravitational waves~\cite{Arzoumanian:2018saf}, the European Pulsar Timing Array~\cite{Lentati:2015qwp}, and the Parkes Pulsar Timing Array~\cite{Hobbs:2013aka,Manchester:2012za}.
The first IPTA data release contains 49 millisecond pulsars that have been observed for 5-30 years with $\mu$s precision~\cite{Verbiest:2016vem}.
Future experiments such as the Square Kilometre Array (SKA) will achieve a much greater sensitivity over current PTAs by finding many new stable millisecond pulsars with better timing precision~\cite{Lazio:2013mea}.
While one of the main goals of PTAs is to detect the stochastic GW background, expected to arise primarily from binaries of supermassive black holes, PTAs are also sensitive to the time-varying gravitational potentials that arise in certain models of DM.

\textit{Substructure and compact objects.}
Massive objects, such as DM subhalos, UCMHs, and PBHs, passing near the Earth-pulsar system can cause a shift in the expected time of arrival of the pulses through two possible mechanisms: the Shapiro time delay~\cite{Siegel:2007fz} and the Doppler effect~\cite{Seto:2007kj}.
The Shapiro delay occurs when the travel time of a pulse is altered due to the presence of a gravitational potential of a transiting object~\cite{Siegel:2007fz,Baghram:2011is,Kashiyama:2012qz,Clark:2015sha,Schutz:2016khr,Dror:2019twh}, while the Doppler effect arises from the object accelerating the Earth or pulsar as it passes by~\cite{Seto:2007kj,Baghram:2011is,Dror:2019twh}.
These two detection strategies cover a wide range of object masses, from $10^{-12}~M_\odot$ to above $100~M_\odot$, complementing lensing searches at low masses and LIGO and LISA searches at high masses.

\textit{Fuzzy DM.}
Another possibility is DM as an ultralight scalar field.
For particles of mass $m$ and velocity $v$, the corresponding de~Broglie wavelength is
\begin{equation}
  \lambda_{\rm dB} \approx 600~\mathrm{pc} \left(\frac{10^{-23}~\mathrm{eV}}{m}\right) \left(\frac{10^{-3}}{v}\right) .
\end{equation}
The wave nature of the DM stabilizes it from collapse on scales of $\lambda_{\rm dB}$, smoothing out inhomogeneities on smaller scales and thereby suppressing structure~\cite{Hu:2000ke}.
For DM within the Galaxy, the scalar field behaves classically.
Its pressure oscillates with an angular frequency $\omega\approx 2m$ and induces oscillations in the gravitational potential~\cite{Khmelnitsky:2013lxt}.
DM masses of $m \sim 10^{-23}$~eV are particularly interesting, since the oscillation frequency $f \sim 5\times 10^{-8} (m/10^{-22}\,\mathrm{eV})$~Hz is in the sensitivity range of PTAs.
Current limits constrains the ultralight DM density to be below $6~\mathrm{GeV}/\mathrm{cm}^3$ for masses $m \lesssim 10^{-23}$~eV~\cite{Porayko:2018sfa}.

\section{Non-perturbative DM dynamics}
\label{sec:NP}

Non-perturbative production and dynamics of DM can give rise to an observable stochastic GW background. In this section we consider two main possibilities: GWs sourced by the breakdown of coherent oscillations of a scalar DM candidate, and GWs sourced by the non-perturbative production of DM.
In each of these scenarios, further dedicated studies will be necessary to pin down the exact relationship between the GW spectrum and the mass and dynamics of the particular DM candidate in question. Nevertheless, the detection of a signal will be a clear sign of new physics.

Homogeneous, oscillating scalar fields provide attractive DM candidates. In the absence of self-interactions, their energy density clusters gravitationally on cosmological time-scales, essentially behaving as CDM on scales larger than the de Broglie scale \cite{Hu:2000ke}. However, when such fields have significant attractive self-interactions, their homogeneous oscillations are unstable to spatial perturbations, leading to self-resonance (parametric and tachyonic), rapid fragmentation, and spatial clustering in the condensate (including formation of solitonic configurations) \cite{Khlopov:1985jw,Amin:2011hj,Lozanov:2017hjm,Amin:2019ums}. Such rapid fragmentation can source a stochastic background of GWs \cite{Kusenko:2008zm,Kitajima:2018zco,Lozanov:2019ylm}. 

Bosonic DM can also be efficiently produced as a daughter field through non-perturbative mechanisms akin to preheating in the early universe (for a review, see \cite{Amin:2014eta}). 
For example, such mechanisms have been proposed as an efficient means of producing cold vector DM as light as $10^{-20}\,\eV$ \cite{Co:2018lka,Bastero-Gil:2018uel,Agrawal:2018vin,Dror:2019twh}.  Such exponential particle production and ensuing nonlinear dynamics can again source a stochastic gravitational wave background due to time dependent anisotropies in the energy-momentum tensor \cite{Khlebnikov:1997di,Easther:2006gt,Dufaux:2007pt}. 
A Chern-Simons coupling between an axion and a dark-photon field strength tensor of the form $(a/f_a) X^{\mu \nu} \tilde{X}_{\mu \nu}$, where $f_a$ is the axion decay constant, causes a tachyonic instability in one of the dark-photon helicities. This leads to a spectrum of chiral GWs where the amplitude is controlled by $f_a/m_\mathrm{pl}$, while the frequency is determined by the axion mass. In particular, the QCD axion can be probed by PTAs if the Chern-Simons coupling is large enough \cite{Machado:2018nqk}.

Nonlinear dynamics of energy densities resulting from the fragmentation of a condensate and production of daughter fields can lead to a detectable stochastic background only if (i) a substantial fraction of the total energy density participates in the production of GWs; (ii) it does so at sufficiently late times (low energies); and (iii) the fragmentation scale is not too small compared to the size of the horizon at the time of production  \cite{Kitajima:2018zco, Amin:2019qrx, Amin:2014eta, Caprini:2018mtu}. For fields that eventually constitute all of the DM in the late Universe, however, it is non-trivial to get their energy fraction to be significant in the early Universe. 

\section{Phase Transitions}
\label{sec:PT}

Phase transitions (PTs) occur when the vacuum state of a theory changes, for example, when a symmetry breaks spontaneously. Phase transitions that feature a discontinuity in the first derivative of the free energy are first-order and inhomogeneous. Bubbles of the new vacuum nucleate in a background of the old vacuum, and as the new vacuum is energetically favored, they expand. Gravitational wave radiation is associated with the collisions of bubbles, as well as the acoustic waves and turbulence in the plasma coupled to the bubble wall. 

The GW spectrum from a first order phase transition (FOPT) is expected to follow a broken power-law, which peaks at a frequency roughly set by the inverse size of the bubbles at collision redshifted to the present time, $f_{\rm peak} \sim R_*^{-1} (a_*/a_0)$ \cite{Hogan:1986qda,Kosowsky:1992rz,Grojean:2006bp,Caprini:2015zlo,Weir:2017wfa,Hindmarsh:2017gnf,Cutting:2018tjt}. 
For transitions which occur during radiation-domination, such as weakly-first-order phase transitions, the peak frequency is predominantly set by the nucleation temperature.
PTAs 
are sensitive to $T_N\sim 10^{-6}-10^{-4}$ GeV, space-based interferometers such as LISA 
are sensitive to PTs around the EW-scale ($T_N\sim 10^{-1}-10^3$ GeV), and ground-based interferometers 
are sensitive to PTs at higher scales $T_N \sim 10^5$ GeV \cite{Croon:2018kqn}. 

Because the Standard Model (SM) does not feature a first-order phase transition~\cite{Kajantie:1996mn,Kajantie:1996qd}, an observed GW background of this kind would point uniquely to new physics (for examples see Refs.~\cite{Madge:2018gfl,Chala:2016ykx}). An observable background also implies that at least an $\mathcal{O}(10^{-2}-10^{-1})$ fraction of the energy density of the Universe was coupled to the order parameter. In this section we review scenarios in which dark matter or a hidden sector generates a FOPT. 
%
%

\subsection{The electroweak phase transition}
A strongly first-order electroweak phase transition is a necessary ingredient for electroweak baryogenesis. As such, there are a variety of well-studied mechanisms for generating such a phase transition. 
For example, it is well known that the addition of a scalar singlet to the SM can promote the SM electroweak phase transition from second-order to first-order by providing an additional cubic term to the effective potential (see e.g.~\cite{Espinosa:2011ax,Cline:2012hg,Curtin:2014jma}). 
If this singlet has an additional $\mathbf{Z}_2$-symmetry, it can also serve as a DM candidate~\cite{Cline:2013gha}, although this condition necessitates that the singlet is a sub-dominant component of the DM relic abundance \cite{Beniwal:2017eik}. 
A singlet scalar can generate Majorana neutrino mass and produce sterile neutrinos with the correct DM relic abundance~\cite{Kusenko:2006rh,Petraki:2007gq,Bian:2018mkl}. 

Models in which new scalars are charged under the SM gauge groups can also impact the electroweak phase transition; the simplest realization of this type of scenario is the addition of a second scalar doublet to the SM Higgs sector, known as the two-Higgs-doublet model (2HDM). There are many variations of 2HDMs, several of which contain DM candidates~\cite{Branco:2011iw} and can lead to a sizable gravitational wave signature~\cite{Kakizaki:2015wua}. Thermal loops of bosons from beyond-Standard-Model (BSM) theories are also a source for a large cubic term in the finite temperature effective potential, and occur in the minimal supersymmetric extension of the SM (MSSM) through the stop squark loop~\cite{Kuzmin:1985mm}, provided that the mass of the lightest stop is below that of the top quark. However, this minimal scenario is in tension with LHC data. Instead, one can consider singlet extensions of the MSSM, for which there are still viable regions of parameter space in which strong electroweak phase transitions are allowed~\cite{Kozaczuk:2014kva}. Finally, the electroweak phase transition can be modified to feature a nearly-conformal potential, inducing a very strong first-order electroweak phase transition. Such possibilities include which also include a dark matter state are composite or Randall-Sundrum models \cite{Randall:2006py,Nardini:2007me,Bruggisser:2018mrt} and extended gauge sector models \cite{Hambye:2013sna,YaserAyazi:2019caf}.
The key point is that for typical electroweak phase transitions, the relevant dimensionful parameters are near the electroweak scale, and thus could potentially be probed by space-based interferometers like LISA.  

\subsection{Phase transitions in hidden sectors}
There are a variety of dynamics associated with a FOPT in a hidden sector that can directly affect the production of the observed DM relic abundance. The first, most widely explored opportunity, is a dark Higgs generating masses in the hidden sector. In this context, the phase transition is not necessarily first-order, nor does it provide direct information into the DM micro-physics. However, models featuring a large number of either gauge bosons or scalars thermally induce sizable potential barriers yielding strong first-order phase transitions \cite{Schwaller:2015tja,Croon:2018erz}. The gauge structure of such models, as well as any additional field content of the theory, all lead to observable deviations in the GW signal \cite{Croon:2018erz}. Note that producing an observable GW signal requires that the hidden sector is at least partially thermalized, so BBN and $\Delta N_\text{eff}$ constraints place strong lower bounds on the scale at which the phase transition can occur \cite{Breitbach:2018ddu,Fairbairn:2019xog}.

Another distinct opportunity relies on phase transitions that occur at a much lower temperature than the critical temperature---a phenomenon termed {\it supercooling}. Given sufficient supercooling, the phase transition drives a period of inflation which dilutes the DM relic density, followed by insufficient reheating to re-thermalize the DM \cite{Konstandin:2011dr,Baratella:2018pxi,Hambye:2018qjv}. For example, the supercooling of the electroweak phase transition down to QCD temperatures leads to sizable GW signatures and affects the abundance of QCD axion DM~\cite{Servant:2014bla,vonHarling:2017yew,Baratella:2018pxi}.
Alternatively, the collision of sufficiently energetic bubble walls can lead to non-thermal production of DM \cite{Chung:1998ua,Konstandin:2011ds,Falkowski:2012fb,Katz:2016adq}. In the latter case, DM significantly heavier than the scale of the electroweak phase transition ($M_\text{DM} \lesssim 10^8~\GeV)$ can be produced. 

\subsection{Hidden sector confinement}
If the dark sector features a gauge coupling that grows large in the IR, i.e.~dark QCD, confinement of dark quarks or gauge bosons will occur. An analytic argument by Pisarski and Wilczek in 1983 \cite{Pisarski:1983ms} determines the confining phase transition to be first order for $N_F\geq3$ light quark flavors at the time of confinement.\footnote{Although the argument relies on an expansion which breaks down for thermal phase transitions, the result is commonly accepted, and demonstrated on the lattice for $N_F = 6$ \cite{Iwasaki:1995ij}.}

Hidden sector confinement has been studied recently in the context of dark quark nuggets \cite{Bai:2018dxf}, as well as solutions to the strong CP problem \cite{Helmboldt:2019pan,Croon:2019iuh}.
Models of confinement invariably predict states around the confinement scale, as well as lighter pions. Therefore, a GW spectrum from a confining phase transition could motivate a collider search for states charged under dark QCD. 

Another potential avenue for studies of DM utilizes the accumulation of DM in front of the bubble wall. As the bubble walls collide, the DM becomes trapped, leading to the formation of large bound states, which can have implications for a number of search strategies such as microlensing and gravitational wave detection \cite{Witten:1984rs,Bai:2018vik,Bai:2018dxf}.


\def\arraystretch{1.2}
\begin{figure*}[tbh!]
\centering
%
\begin{minipage}{0.745\linewidth}
   \includegraphics[width=\linewidth]{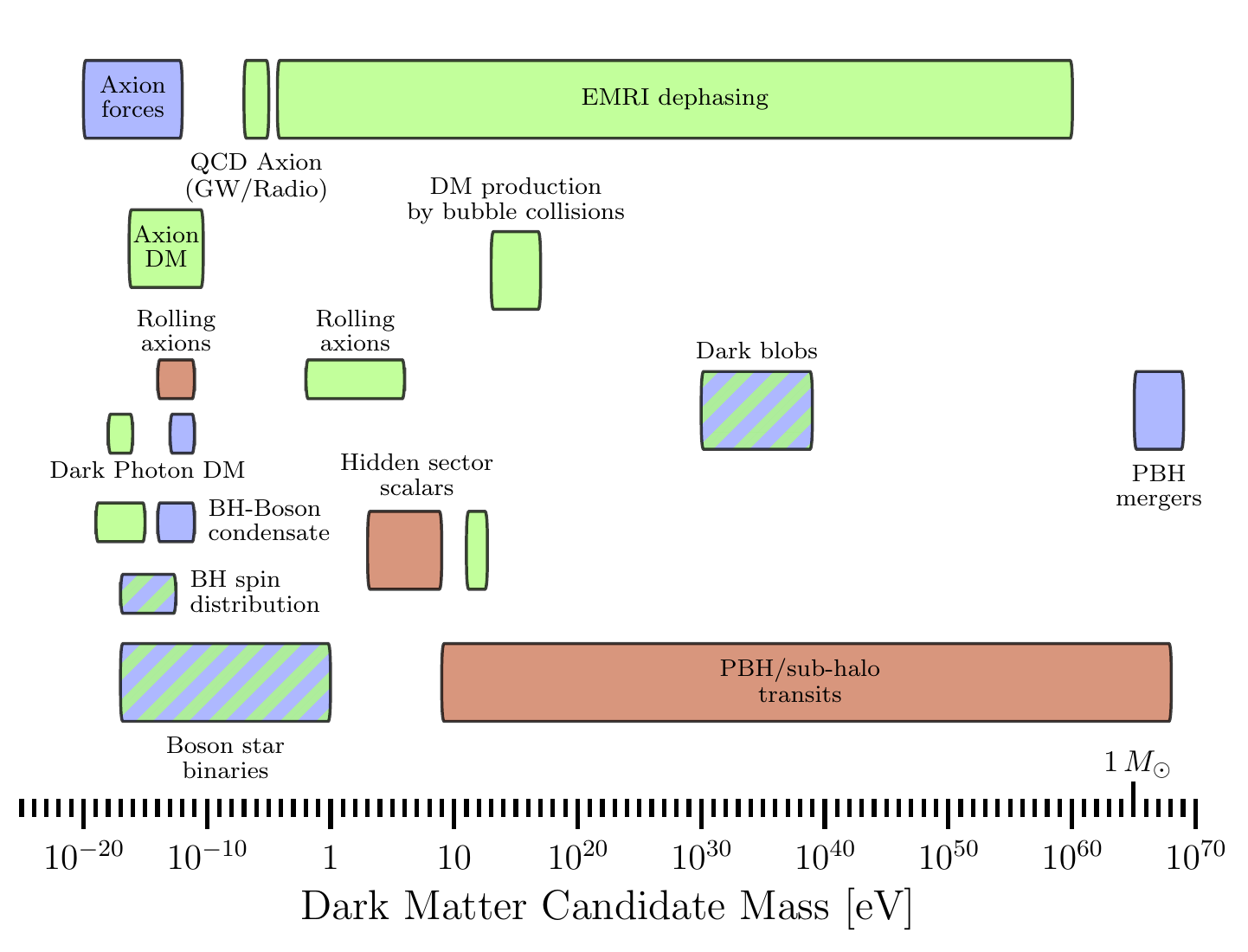}
\end{minipage}
\begin{minipage}[c][][t]{0.245\linewidth}
\includegraphics[width=0.9\linewidth]{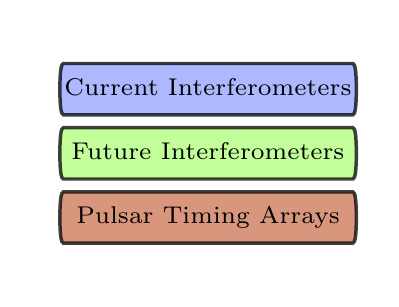}
\begingroup
\fontsize{8}{10}\selectfont
            \begin{tabularx}{\textwidth}{l} 
                \toprule
                {\small Constraints}\\
                \colrule
                Axion forces \cite{Huang:2018pbu} \\
                QCD Axion (GW/Radio)  \cite{Edwards:2019tzf}\\
                EMRI dephasing  \cite{Eda:2013gg,Eda:2014kra} \\
                Axion DM \cite{Nagano:2019rbw}\\
                Bubble collision DM \cite{Falkowski:2012fb}\\
                Rolling axions  \cite{Machado:2018nqk}\\
                Dark Photon DM \cite{Pierce:2018xmy}\\
                Dark blobs \cite{Grabowska:2018lnd}\\
                PBH mergers  \cite{Ali-Haimoud:2017rtz,Kavanagh:2018ggo} \\
                BH-boson condensate  \cite{Arvanitaki:2014wva,Brito:2017zvb}\\
                Hidden-sector scalars  \cite{Croon:2018erz,Breitbach:2018ddu}\\ 
                BH spin distribution  \cite{Arvanitaki:2014wva,Brito:2017zvb}\\
                Boson stars \cite{Palenzuela:2017kcg,Helfer:2018vtq,Croon:2018ftb,Guo:2019sns}\\
                PBH/sub-halo transits  \cite{Schutz:2016khr,Dror:2019twh}\\
                 \botrule
            \end{tabularx}
\endgroup
\end{minipage}
       \caption{\textbf{Summary of possible gravitational wave constraints on dark matter.}  We show the ranges of DM mass which are covered by potential constraints from gravitational waves, using current interferometers (blue), future interferometers (green) and Pulsar Timing Arrays (red). 
       Some representative references for each constraint are shown in the table on the right. See the main text for further details and references.}
    \label{fig2}
\end{figure*}

\section{Discussion and conclusions}

We summarise current and future avenues for constraining dark matter using gravitational waves in Fig.~\ref{fig2}. The range of possible constraints is impressive, as it spans a wide range of detection techniques and DM candidates and almost 90 orders of magnitude in DM candidate mass. 

We also remark that GWs allow us to constrain alternative theories of gravity, some of which have been proposed as alternatives to particle DM. The coincident observation of electromagnetic radiation and GWs from the merger of two
neutron stars  \cite{TheLIGOScientific:2017qsa} has for instance already provided stringent constraints on theories of modified gravity in which photons travel on different geodesics with respect to GWs \cite{Boran:2017rdn,Sakstein:2017xjx,Wang:2017rpx}.

The connection between DM and GWs remains, however, largely unexplored, and much remains to be done to improve existing constraints and to assess in a more complete and robust fashion the prospects for identifying DM using future experiments. We highlight here a list of challenges and open questions for each of the research areas discussed above:

\begin{itemize}
\item {\bf Primordial black holes.}  What is the initial distribution of PBHs and how does it affect the GW signal and other probes? Can observations help us discriminate between different PBH production scenarios? How can we probe sub-lunar mass PBHs constituting DM? What type of radiation is emitted by accreting PBHs? Are PBHs the seeds of supermassive BHs? How do constraints on PBHs change in the presence of other DM candidates?
\item {\bf Environmental effects.} What is the effect of an inspiraling object on the DM halo surrounding an intermediate-mass or supermassive black hole? Is the halo disrupted, or destroyed? How common and robust is such a DM overdensity? And would its presence be detectable through a dephasing in the gravitational waveform?
\item {\bf Exotic Compact Objects.} What is their formation history and their distribution in the Galaxy? What is the impact on the number of predicted events and on the amplitude of the stochastic gravitational wave background they induce?
\item {\bf Direct detection with GW experiments.} What modifications of current and upcoming interferometers would maximize the sensitivity to DM models without jeopardizing gravitational wave searches? What would we learn about the nature of DM from the detection of subhalos with PTAs?
\item {\bf Non-perturbative DM dynamics.} Thorough analyses have been performed mostly in the context of the very early Universe, where the typical GW frequencies are much higher than any current experiment. 
Is it possible to get a measurable GW signal in the case of fragmentation of a coherent field or non-perturbative particle production in the relatively late Universe?
\item {\bf Phase transitions.} A broken power law in the stochastic background would imply new physics comprising a large fraction of the energy density at the time of the transition. Can we break the degeneracy among models predicting this feature, e.g.\ by searching for new states with masses at scales similar to that of the nucleation temperature of the phase transition?
\end{itemize}

In conclusion, we believe that the time is ripe for investigating the many connections between GWs and DM, and we encourage a strong community effort to further explore the interface between these two exciting fields of research. 

\begin{acknowledgments}
We thank Robert Caldwell, Vitor Cardoso, Jeff Dror, Thomas Edwards, Alex Kusenko, Ranjan Laha, David Morrissey, Rafael Porto, Nirmal Raj, Pedro Schwaller, Géraldine Servant, Chen Sun, and David Weir for  useful comments and suggestions. This work was initiated and performed in part at the Aspen Center for Physics, which is supported by National Science Foundation grant PHY-1607611. 
\end{acknowledgments}

\bibliography{refs}

\end{document}